%% file: 19SPAWC_SSCC.tex
\documentclass[conference]{IEEEtran}
\IEEEoverridecommandlockouts

 \usepackage[T1]{fontenc}
\usepackage{amsmath,amsfonts} %citesort removed
\usepackage{amssymb, amsthm}
\usepackage{graphicx}
\usepackage{subfigure}
\usepackage{booktabs}
\usepackage{multirow}
\usepackage{mathrsfs}
\usepackage{units}
\usepackage{cancel}
\usepackage{upgreek}
\usepackage{color}
\usepackage{microtype}
\usepackage{balance}
\usepackage{paralist}
\usepackage{enumitem}
\usepackage{afterpage}
\usepackage{wrapfig}
\usepackage{balance}
\usepackage{cite}

\input{vmr-symbols-vecbold}
\input{standard-macros}

\input{defs}

\renewcommand{\bml}{\ensuremath{\boldsymbol \ell}}

\title{Improving Channel Charting with Representation-Constrained Autoencoders}

\author{\IEEEauthorblockN{Pengzhi Huang$^1$,  Oscar Casta\~neda$^1$, Emre G\"{o}n\"{u}lta\c{s}$^1$, Sa\"id Medjkouh$^1$,  \\ Olav Tirkkonen$^2$, Tom Goldstein$^3$, and Christoph Studer$^1$} \\[-0.25cm]
\IEEEauthorblockA{
\small $^1$\textit{School of Electrical and Computer Engineering, Cornell University, Ithaca, NY; email:  {studer@cornell.edu}} \\
$^2$\textit{School of Electrical Engineering, Aalto University, Finland; e-mail: {olav.tirkkonen@aalto.fi}}\\
$^3$\textit{Department of Computer Science, University of Maryland, College Park, MD; e-mail: {tomg@cs.umd.edu}}\thanks{The work of PH, OC, EG, SM, and CS was supported by Xilinx, Inc. and by the US NSF grants  ECCS-1408006, CCF-1535897,  CCF-1652065, CNS-1717559, and ECCS-1824379. 
The work of TG was supported by the US NSF under grant CCF-1535902 and by the US Office of Naval Research grant \mbox{N00014-17-1-2078}. The work of OT was funded in part by the Academy of Finland (grant 319484). 
}
}
}

\begin{document}

\maketitle

\begin{abstract}
Channel charting (CC) has been proposed recently to enable logical positioning of user equipments (UEs) in the neighborhood of a multi-antenna base-station solely from channel-state information (CSI). CC relies on dimensionality reduction of high-dimensional CSI features in order to construct a channel chart that captures spatial and radio geometries so that UEs close in space are close in the channel chart. 
In this paper, we demonstrate that autoencoder (AE)-based CC can be augmented with side information that is obtained during the CSI acquisition process. 
More specifically, we propose to include pairwise representation constraints into AEs with the goal of improving the quality of the learned channel charts.
We show that such representation-constrained AEs recover the global geometry of the learned channel charts, which enables CC to perform approximate positioning without global navigation satellite systems or supervised learning methods that rely on extensive and expensive measurement campaigns. 
\end{abstract}

 %%%
\section{Introduction}
Autoencoders (AEs) are single- or multi-layer neural networks widely used for dimensionality-reduction tasks \cite{HintonSalakhutdinov2006b,van2009dimensionality, bengio, pmlr-v27-baldi12a}. AEs learn low-dimensional representations (embeddings) of a given high-dimensional dataset and have been shown to accurately preserve spatial relationships in both high- and low-dimensional space for a broad range of synthetic and real-world datasets \cite{van2009dimensionality}.
With the success of deep neural networks, AEs are also gaining increased attention for unsupervised learning tasks~\cite{7286736}.
Notable application examples of AEs include learning word embeddings \cite{NIPS2013_5021},  image compression \cite{2017arXiv170300395T}, generative models  \cite{journals/corr/abs-1305-6663}, and channel charting \cite{cc_paper,cc_paper_globecom}.
AEs are typically trained in an unsupervised manner, i.e., no labels are used, while potential side information on the training data is routinely ignored or application-specific representation structure is not imposed during training.

AEs that impose structural constraints on the latent variables include sparse AEs \cite{Hinton2006FLA} and variational AEs \cite{2016arXiv160605908D}. Sparse AEs enforce sparsity on the representations, which enables one to learn embeddings with low effective dimensionality. 
Variational AEs learn an embedding drawn from a distribution that represents the high-dimensional input \cite{2016arXiv160605908D}; such AEs have been shown to be able to generate complex data, such as handwritten digits \cite{2013arXiv1312}, faces \cite{2015arXiv150303167K}, or physical scenes \cite{2015arXiv150303167K}.

\subsection{Representation-Constrained Autoencoders for Positioning}
A range of  dimensionality-reduction applications provide valuable side information that can be imposed on the low-dimensional representations. Such side information may stem either from the dataset itself or from the way data was collected. 
One example arises when data is acquired over time, where it may be natural to  enforce constraints \emph{between} representations by exploiting the fact that, for temporally correlated datapoints, the associated low-dimensional representations should be similar. 
Another example arises when a subset of the representations are known a-priori, e.g., when a small part of the training data has been annotated. The obtained information can then be translated into representation constraints, which leads to semi-supervised training of AEs. 

{Representation constraints} are important for positioning users in wireless systems using \emph{channel charting} (CC) \cite{cc_paper,cc_paper_globecom}. CC measures high-dimensional channel-state information (CSI) of user equipments (UEs) transmitting data to an access point or cell tower. By collecting CSI at multiple spatial locations over time, one can train an AE for which the low-dimensional representations reflect relative UE positions. While the original CC method~\cite{cc_paper} enables relative localization without access to global navigation satellite systems (GNSS) and without dedicated measurement campaigns \cite{4343996,7239531}, valuable side information should not be ignored when available.
As UEs move with finite velocity, one could consider this information when training an AE to ensure that temporally correlated datapoints are nearby in the representation space. One could also imagine that certain points in space with known location (e.g., a coffee shop) can be associated with measured CSI; this helps to pin down a subset of spatial locations in the representation space, which enables absolute positioning. 
Put simply, enforcing constraints \emph{between} representations may improve the efficacy of AE-based CC.

\subsection{Contributions}
This paper investigates representation constraints for AEs and provides a framework for including these during training. 
We propose constraints on pairs of representations in which either the absolute or relative distance among (a subset of) representations is enforced. 
We formulate these constraints as nonconvex regularizers, which can easily be included into well-established deep-learning frameworks. 
We highlight the efficacy of representation constraints for the application of CC-based positioning in wireless systems: in  particular, we demonstrate that by combining partially-annotated locations with temporal UE constraints, the positioning performance of CC \cite{cc_paper} can be improved significantly.

\subsection{Relevant Prior Art}
Autoencoders provide excellent dimensionality-reduction performance with real-world datasets \cite{van2009dimensionality}. Despite their success, AEs do not aim at preserving geometric properties on the representations (such as distances between points); this is in stark contrast to, e.g., Sammon's mapping \cite{sammon} or multidimensional scaling \cite{davison1991multidimensional}. 
Furthermore, AEs commonly ignore side information that stems from the application at hand. We will extend AEs with \emph{pairwise representation constraints} that can improve performance of dimensionality reduction tasks.

The efficacy of deep learning has been explored recently for wireless positioning \cite{confi,8292280,2018arXiv180404826A}. The methods in these papers rely on extensive measurement campaigns and require CSI measurements annotated with exact position information.
To avoid the drawback of such supervised methods, CC, as put forward in~\cite{cc_paper}, uses dimensionality reduction to extract {\em relative} position information of users without the necessity of costly measurement campaigns. 
CC exploits the fact that CSI is high-dimensional, but strongly depends on UE position, which is low-dimensional. Dimensionality reduction applied to CSI measurements learns a \emph{channel chart}, in which nearby points represent nearby locations in true space---exact position information is not available. 
However, CC naturally provides representation constraints originating from the acquisition process---we will include such constraints into AEs to significantly improve CC.

\section{Representation-Constrained Autoencoders}
We briefly summarize the basics of AEs and then introduce the concept of pairwise representation constraints.

\subsection{Autoencoders in a Nutshell}
AEs take a high-dimensional dataset consisting of $N$ datapoints (vectors) $\bmx_n\in\reals^D$, $n=1,\ldots,N$, of dimension $D$, and learn two functions: the encoder $f^e:\reals^{D}\to \reals^{D'}$ and the decoder $f^d:\reals^{D'}\to \reals^{D}$. The encoder maps datapoints onto low-dimensional \emph{representations} $\bmy_n\in\reals^{D'}$, $n=1,\ldots,N$, where $D'\ll D$ is the dimension of the representation, and the decoder maps representations back to datapoints:
\begin{align*}
\bmy_n=f^e(\bmx_n) \quad \text{and} \quad \bmx_n=f^d(\bmy_n), \quad n=1,\ldots,N.
\end{align*}
The encoder and decoder functions of AEs are implemented as multilayer (shallow or deep) feed-forward neural networks that are trained to minimize the mean-square error (MSE) between the input and the output of the network, specifically:
\begin{align} \label{eq:approximationerror}
L(\setW^e,\setW^d) = \frac{1}{N}\sum_{n=1}^N\|\bmx_n - f^d(f^e(\bmx_n))\|^2.
\end{align}
The parameters to be learned from $\{\bmx_n\}_{n=1}^N$ are the weights and bias terms contained in the sets $\setW^e$ and $\setW^d$, which define the encoder and decoder neural networks, respectively. 

The $D'$-dimensional output of the encoder $f^e$ is typically of lower dimension than the intrinsic dimension of the manifold embedding the inputs $\bmx_i$ in $D$ dimensions. 
Hence, we have that $\bmx_n\approx f^d(f^e(\bmx_n))$, $n=1,\ldots,N$, unless the dataset $\{\bmx_n\}_{n=1}^N$ was $D'$-dimensional and we were able to learn the underlying structure.
Nevertheless, AEs often find low-dimensional representations $\{\bmy_n\}_{n=1}^N$ with small loss that capture the intrinsic dimensionality of the input datapoints.

\setlength{\textfloatsep}{5pt}% Remove \textfloatsep
\begin{table*}[tp]
\centering
\caption{Summary of proposed representation constraints for AEs (known quantities are underlined).}
\label{tbl:constraintssummary}
\begin{tabular}{@{}lll@{}}
\toprule
\bf Name & \bf Constraint & \bf Regularizer  \\
\midrule
Fixed absolute distance (FAD) & $\|\bmy_i-\underline{\bmy}_j\| = \underline{d}_{i,j} $  & $(\|\bmy_i-\underline{\bmy}_j\|-\underline{d}_{i,j})^2$  \\
Fixed relative distance (FRD)  & $\|\bmy_i-\bmy_j\| = \underline{d}_{i,j} $  & $(\|\bmy_i-\bmy_j\|-\underline{d}_{i,j})^2$  \\
Maximum absolute distance (MAD) & $\|\bmy_i-\underline{\bmy}_j\| \leq \underline{d}_{i,j} $ & $\max\{\|\bmy_i-\underline{\bmy}_j\|-\underline{d}_{i,j},0\}^2$ \\
Maximum relative distance (MRD) & $\|\bmy_i-\bmy_j\| \leq \underline{d}_{i,j} $ &  $\max\{\|\bmy_i-\bmy_j\|-\underline{d}_{i,j},0\}^2$ \\
\bottomrule
\vspace{-0.6cm}
\end{tabular}
\end{table*}

%%%
\subsection{Pairwise Representation Constraints}
\label{sec:repcons}
In \fref{tbl:constraintssummary}, we propose four distinct pairwise representation constraints, where the underlined quantities represent {\em constant} scalars or vectors that are known a-priori and used during AE training; non-underlined quantities are optimization variables.

%%%
\subsubsection{Fixed Distance Constraints}
\label{sec:FADconstraints}
The fixed absolute distance (FAD) and fixed relative distance (FRD) constraints enforce a known distance $\underline{d}_{i,j}$ on a pair of representations according to $\|\bmy_i-\bmy_j\|=\underline{d}_{i,j}$. The difference between FAD and FRD is that, for FAD, one of the two representations, e.g., $\underline{\bmy}_{j}$, is a constant known prior to AE learning; for FRD, both representations $\bmy_i$ and $\bmy_j$ are optimization variables.
To facilitate the inclusion of these constraints in deep learning frameworks, we propose to use regularizers (see  \fref{tbl:constraintssummary}) for which generalized gradients exist. 
Concretely, the generalized gradient of the FRD constraints with respect to the representation~$\bmy_i$ is  
\begin{align} \label{eq:FADFRDgradients}
    \nabla_{\bmy_i}(\|\bmy_i\!-\!{\bmy}_j\|\!-\!\underline{d}_{i,j})^2 = 2(\|\bmy_i\!-\!{\bmy}_j\|\!-\!\underline{d}_{i,j}) \frac{\bmy_i\!-\!\bmy_j}{\|\bmy_i\!-\!\bmy_j\|},
\end{align}
where for the FAD constraint the representation~$\bmy_j$ is known a priori, i.e., $\bmy_j=\underline{\bmy}_j$.
If $\underline{d}_{i,j}=0$, then the FRD regularizer promotes equality among $\bmy_i$ and~$\bmy_j$, whereas the FAD regularizer will learn a representation~$\bmy_i$ that is close to the constant vector $\underline{\bmy}_j$.
Intuitively, the FAD constraint for $\underline{d}_{i,j}=0$ acts as a semi-supervised extension in which a subset of representations are known a-priori. 

%%%
\subsubsection{Maximum Distance Constraints}
The maximum absolute distance (MAD) and maximum relative distance (MRD) constraints enforce a maximum a-priori known distance $\underline{d}_{i,j}$ between a pair of representations according to $\|\bmy_i-\bmy_j\|\leq\underline{d}_{i,j}$. For MAD, one of the two vectors in the constraint, e.g., $\underline{\bmy}_{j}$, is known a-priori; for MRD, both representations are learned. 
We include these constraints as regularizers (see  \fref{tbl:constraintssummary}) with the generalized gradient 
\begin{align} \label{eq:MADMRDgradients}
    &\nabla_{\bmy_i}\max\{\|\bmy_i-\bmy_j\|-\underline{d}_{i,j},0\}^2 \notag  \\
     & \qquad  = 2\max\{\|\bmy_i-\bmy_j\|-\underline{d}_{i,j},0\} \frac{\bmy_i-\bmy_j}{\|\bmy_i-\bmy_j\|},
\end{align}
where $\bmy_j=\underline{\bmy}_j$ is known for MAD. 
Note that if $\underline{d}_{i,j}=0$, then FAD is equivalent to MAD and FRD is equivalent to MRD.

%%%
\subsubsection{Practical Considerations}
We implemented a stochastic optimizer to minimize the sum of the AE fidelity term \fref{eq:approximationerror} and the regularized constraint penalties using the Keras and TensorFlow frameworks \cite{tensorflow2015-whitepaper,chollet2015keras}.  Because penalty terms may represent pairwise constraints that involve two data points, the stochastic approximation of the regularizers was formed by randomly sampling constraints rather than datapoints.

\subsection{Performance Metrics}
In order to measure the performance of dimensionality reduction, we use two standard metrics for the local neighborhood-preserving performance: \emph{trustworthiness} (TW) and \emph{continuity} (CT). 
TW measures whether mapping high-dimensional datapoints to the representation space introduces new (false) neighbors. TW is defined as
\begin{align*}
    \textit{TW}(K) = \textstyle 1- \frac{2}{NK(2N-3K-1)}\sum_{i=1}^N \sum_{j\in\setU^K_i}(r(i,j)-K),
\end{align*}
where $r(i,j)$ represents the rank of the representation $\bmy_i$ among the pairwise distances between the other representations. The set $\setU^K_i$ contains the points that are among the $K$ nearest neighbors in representation space, but not in the high-dimensional space. 
CT measures if similar datapoints in original space remain similar in the representation space, and is defined as
\begin{align*}
    \textit{CT}(K) = \textstyle  1-\frac{2}{NK(2N-3K-1)}\sum_{i=1}^N \sum_{j\in\setV^K_i} (\hat{r}(i,j)-K), 
\end{align*}
where $\hat{r}(i,j)$ represents the rank of the datapoint $\bmx_i$ among the pairwise distances between the other datapoints. The set $\setV^K_i$ contains the  points that are among the $K$ nearest neighbors in the high-dimensional space, but not in the representation space. 
Both TW and CT have values in the range $[0,1]$ and large values indicate that neighborhoods are better preserved.

Besides measuring the local-neighborhood-preservation properties via TW and CT, we also consider Kruskal's stress (KS) \cite{lee2009dr, shepard1962}, which measures how well the global structure in the high-dimensional dataset $\{\bmx_n\}_{n=1}^N$ is mapped to the low-dimensional embedding  $\{\bmy_n\}_{n=1}^N$. KS is defined as
\begin{align*}
\textit{KS} =  \sqrt{\frac{\sum_{n,m}(\delta_{n,m}-\beta\hat{\delta}_{n,m})^2}{\sum_{n,m}{\delta_{n,m}^2}}},
\end{align*}
where $\delta_{n,m}=\|\bmx_n-\bmx_m\|$, $\hat{\delta}_{n,m}=\|\bmy_n-\bmy_m\|$, and $\beta=\sum_{n,m}\delta_{n,m}\hat{\delta}_{n,m}/\sum_{n,m}\delta_{n,m}^2$ is the optimal distance scaling factor. KS is in the range $[0,1]$ and smaller values indicate that global geometrical structure is preserved better. If $\textit{KS}=0$, then the geometry is perfectly preserved. 

%%%
\section{Channel Charting with Representation-Constrained Autoencoders}
We now augment the original CC framework \cite{cc_paper} with representation constraints that naturally arise from the application. 
We start by outlining the concept of CC and then explain how representation constraints are included. We then demonstrate the efficacy in comparison to the original CC framework.

\subsection{Channel Charting in a Nutshell}
 
CC measures CSI from users at different spatial locations and learns a low-dimensional \emph{channel chart} that preserves \emph{locally} the original spatial geometry. Put simply, users that are physically nearby will be placed nearby in the channel chart and vice versa---global geometry is typically not preserved.
In this framework, high-dimensional features are extracted from CSI, then processed with dimensionality-reduction methods to obtain the low-dimensional channel chart. CC operates in an unsupervised manner, i.e., learning is only based on CSI that is passively collected at an infrastructure base-station (BS), but from multiple user locations in the service area over time.
CC opens up many location-based applications as it provides BS providers with relative user location information without access to GNSS or fingerprinting methods \cite{4343996}. 

The technical concepts behind channel charting are as follows. Suppose that we have $N$ single-antenna users located in real space with coordinates $\bmz_n\in\reals^3$.
If the $n$th user at location $\bmz_n$ is transmitting data to a BS, then the BS first extracts high-dimensional CSI in the form of a high-dimensional vector $\bmh_n\in\complexset^D$, which represents multi-path scattering and path loss of the wireless channel. 
From the CSI vector $\bmh_n$, one can extract features $\bmx_n\in\reals^D$ that represent large-scale fading properties of the wireless channel. 
The main assumption of CC is that large-scale fading properties are mostly static and are strongly tied to user location. Specifically, due to the underlying physics of electromagnetic wave propagation, each CSI feature is a (noisy) function of the user position, a function that represents the effect of the (unknown)  physical environment on the transmitted signal.
One can then learn the channel chart from the set of channel features $\{\bmx_n\}_{n=1}^{N}$ in an unsupervised manner by means of dimensionality-reduction methods. 
If AEs are used in this procedure, the encoder~$f^e$ corresponds to the forward charting function that maps CSI features $\bmx_n$ to relative position information $\bmy_n$ in the representation space. 

Reference \cite{cc_paper} proposed the use of Sammon's mapping (SM) and AEs to learn the channel charts. While SM exhibited good performance, AEs scale well to large problem sizes and provide a parametric mapping that enables one to map new, unseen CSI features to a relative location. 
Despite the advantages of AEs, valuable side information that arises from the application itself has been ignored. 
First, in contrast to SM, conventional AEs do not enforce any geometric structure on their representations. Second, by tracking a user's CSI over time, the corresponding low-dimensional representations that reflect their position should be similar as velocity is finite. 

\subsection{Channel Charting with Representation Constraints}
We propose the inclusion of representation constraints to overcome the limitations of the original CC framework. 
We impose MRD constraints on pairs of representations from a user over time, to ensure nearby representations for nearby spatial locations. We can estimate an upper limit on the maximum distance in the representation space based on the measurement CSI acquisition rate. Note that this information comes directly from the CSI measurement process and the fact that we know how data was collected. 

Furthermore, to enable CC with true positioning capabilities, we unwrap the channel chart using \emph{anchor vectors}, i.e., points in space for which we know both their CSI as well as their true location. One can imagine measuring CSI at a small set of locations when setting up a new BS (e.g., by knowing the precise location of a  fixed access point). 
With this information, we impose FAD representation constraints on the AE  with $\underline{d}_{i,j}=0$ to enforce the exact anchor positions. 
The inclusion of such constraints leads to a semi-supervised version of CC (and AEs in general). In contrast to conventional fingerprinting methods that are fully supervised and require training at wavelength resolution in space, we only require a small number of anchor vectors and use the rest of the (unlabeled) data to improve the localization accuracy of the channel chart.

\subsection{Numerical Results}
%

%\begin{wrapfigure}[24]{r}{0.5\textwidth}
\setlength{\textfloatsep}{5pt}% Remove \textfloatsep
\begin{figure}[tp]  
\centering
%\vspace{-1.0cm}
\includegraphics[width=0.41\columnwidth]{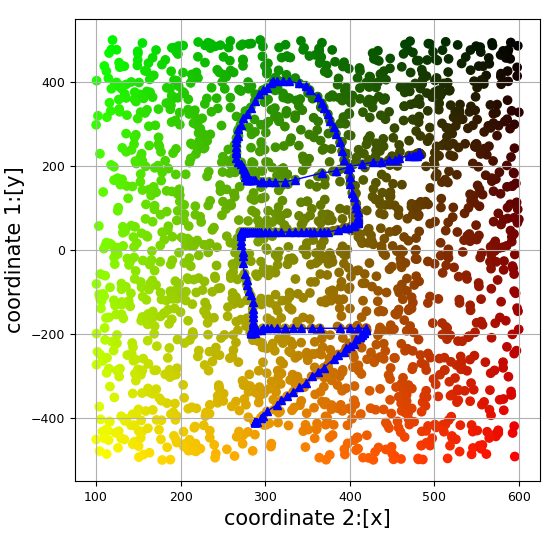}
\vspace{-0.2cm}
\caption{Channel charting scenario as in \cite{cc_paper}. A 32-antenna BS located at the origin extracts 2048 CSI vectors from several UEs in space. Each point represents a UE position; the points on the ``vip'' curve model a user that moves in time. The color gradients and ``vip'' curve enable an easy visual comparison with the learned channel charts shown in \fref{fig:channelcharts}.}
\label{fig:scenario} 
\end{figure}
%\end{wrapfigure}

\subsubsection{Scenario}
\fref{fig:scenario} depicts the scenario.
We measure the CSI of $N=2048$ randomly placed user locations (except for the positions on the ``vip'' curve) within a rectangular area of $1000$\,m $\times$ $500$\,m. We model the acquisition of CSI at 0\,dB SNR. At spatial location $(x,y,z)=(0,0,10)$ meters, we consider a uniform linear BS antenna array with half-wavelength spacing and $32$ antennas. 
We model narrowband data transmission at $2$\,GHz and consider two channel models: (i) Quadriga LoS (Q-LoS; a realistic model for LoS channels that includes scatterers and path loss); and (ii) Quadriga non-LoS (Q-NLoS; a realistic model for channels where there is no direct path between users and the BS antenna array).
For both models, we used the ``Berlin UMa'' scenario, which has been calibrated with real-world measurements~\cite{jaeckel2014quadriga}. 
At the BS side, we extract the same features proposed in \cite{cc_paper}, i.e., we apply feature scaling, convert them to the angular domain, and take the entry-wise absolute value.
The input has $D=32$ real dimensions; the representation dimension is $D'=2$.
We use a similar structure of the AE in \cite{cc_paper}, namely $9$ hidden dense layers: $4$ dense layers for the encoder and another $4$ dense layers for the decoder (the layers consist of $500$, $100$, $50$ and $20$ neurons), and an intermediate layer where we extract the channel chart with 2 neurons and linear activation functions.

\setlength{\textfloatsep}{5pt}% Remove \textfloatsep
\newcommand{\crazyscale}{0.41}
\begin{figure}[t] \label{fig:charts}
\centering
\subfigure[Q-LoS; plain]{\includegraphics[width=\crazyscale\columnwidth]{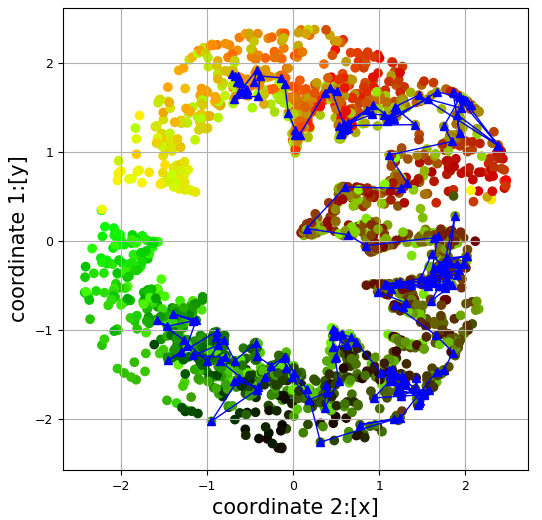}}
\hspace{0.4cm}
\subfigure[Q-NLoS; plain]{\includegraphics[width=\crazyscale\columnwidth]{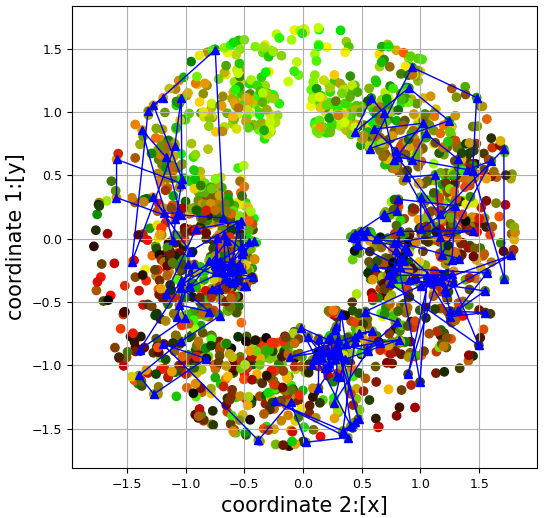}}\\ 
\vspace{-0.3cm}
\subfigure[Q-LoS; FAD]{\includegraphics[width=\crazyscale\columnwidth]{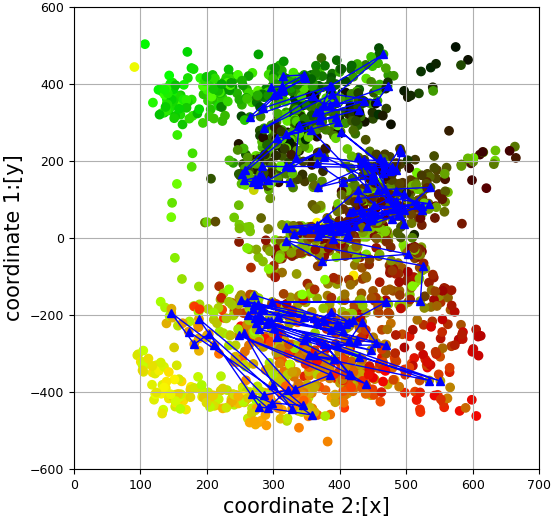}}
\hspace{0.4cm}
\subfigure[Q-NLoS; FAD]{\includegraphics[width=\crazyscale\columnwidth]{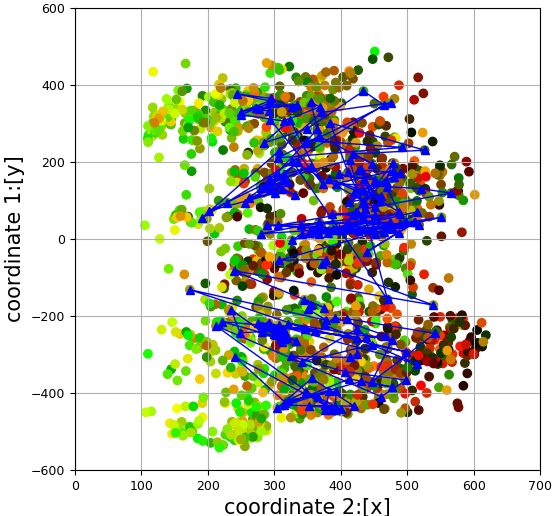}} \\
\vspace{-0.3cm} 
\subfigure[Q-LoS; FAD\&MRD]{\includegraphics[width=\crazyscale\columnwidth]{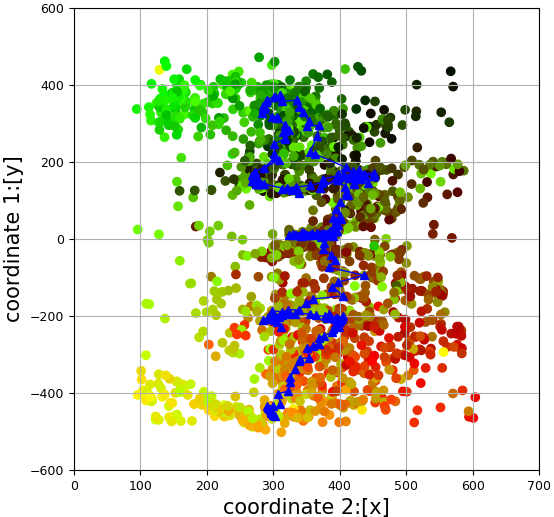}}
\hspace{0.4cm}
\subfigure[Q-NLoS; FAD\&MRD]{\includegraphics[width=\crazyscale\columnwidth]{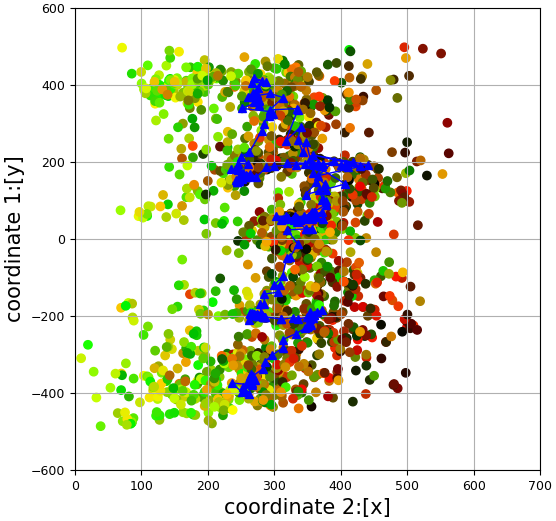}}
\vspace{-0.1cm} 
\caption{Channel charts learned from the scenario in \fref{fig:scenario}. The channel charts of plain AEs represent well the local neighborhood structure, but do not capture the global geometric properties. Imposing FAD and MRD constraints recovers the fine aspects of the geometry, enabling approximate positioning.}
\label{fig:channelcharts}
\end{figure}

\subsubsection{Results}
\fref{fig:channelcharts} shows the channel charts. 
The top row (Figures~\ref{fig:channelcharts}(a)--(b)) shows the results of a ``plain'' AE, i.e., without any representation constraints; these results reproduce those in \cite{cc_paper}.
The middle row (Figures~\ref{fig:channelcharts}(c)--(d)) shows channel charts from  AEs that include FAD constraints,  where we randomly selected $10\%$ of the users to be anchor vectors. Clearly, these FAD constraints unwrap the channel chart and lead to a representation with a distance scale comparable to the original scenario in \fref{fig:scenario}. While the global structure is approximately preserved, the points on the ``vip'' curve are not represented accurately. 
The bottom row (Figures~\ref{fig:channelcharts}(e)--(f)) shows the combination of FAD and MRD constraints in AEs. Concretely, we also enforce the fact that points on the ``vip'' curve model a user's motion and we can impose maximum absolute distance constraints among pairs of representations pertaining to this curve. FAD and MRD combined are able to reproduce the original scenario: It is evident that points in the channel chart approximately represent the true locations. 
We also observe that the propagation conditions of the wireless channel do not substantially affect the performance.

\begin{table}[tp]
    \centering
    \renewcommand{\arraystretch}{1.05}
    \caption{TW, CT, and KS results for channel charting with and without representation constraints.}
    \label{tbl:CCresults}
    \scalebox{0.825}{ 
    \begin{tabular}{@{}ll|ccc|ccc@{}}
    \toprule
    \multicolumn{2}{c}{} & \multicolumn{3}{c}{\bf Q-LoS} & \multicolumn{3}{c}{\bf Q-NLoS} \\
    \midrule
    &  & Plain & FAD & FAD\&MRD & Plain & FAD & FAD\&MRD \\
    \midrule
    \multirow{3}{*}{TW} 
    & $K=1$   &           0.8468 & 0.8516 & \bf0.8576      & 0.8480 & 0.8492  & \bf0.8597  \\
    & $K=51$  &   0.8597 & 0.8570 & \bf 0.8651      & 0.8502 & 0.8560 & \bf 0.8665  \\
    & $K=102$ &  0.8609 & 0.8642 & \textbf{0.8736} & 0.8546 & 0.8626 &\textbf{ 0.8739 } \\
    \midrule
    \multirow{3}{*}{CT} 
    & $K=1$   &  \bf 0.9700 & 0.9195 & 0.9321      & \textbf{0.9281} & 0.8924 &  0.9110   \\
    & $K=51$  &  \bf 0.9440 & 0.9067 & 0.9215      & \bf 0.9168 & 0.8928 & 0.9128 \\
    & $K=102$ &   \bf 0.9358 & 0.9081 & 0.9216      & \bf 0.9155 & 0.8964 & 0.9151  \\
    \midrule
    KS &        &   0.3548 & 0.2652 & \bf 0.2598      & 0.4096 & 0.2749 & \bf 0.2693  \\
    \bottomrule
    \vspace{-0.1cm}
    \end{tabular}
    }
\end{table}

\fref{tbl:CCresults} lists the TW, CT, and KS results for the channel charts shown in \fref{fig:channelcharts}. We see that including representation constraints improves TW while slightly lowering the CT; note that TW and CT are evaluated for $K= 1$, $K=51$ ($2.5$\% of the dataset), and $K=102$ ($5$\% of the dataset). 
Hence, we observe a tradeoff with respect to neighborhood-preserving properties. 
More concretely, an increase in TW means that we are introducing fewer ``fake'' near neighbors; a reduction in CT means that neighborhoods in the original space are not as well preserved in the channel chart as before. 
With respect to the global geometric structure, we see that KS significantly improves for all representation-constrained AEs; this implies that the inclusion of constraints enables us to recover global geometry.
We note this is also visible in \fref{fig:channelcharts}, especially in the AE results that include both FAD constraints (anchor vectors) and MRD constraints (to enforce continuity of a user's motion).
Once again, we see that the propagation conditions do not substantially affect the performance.

%%%
\section{Conclusions}

We have shown how to incorporate pairwise representation constraints into autoencoders (AEs). 
To demonstrate effectiveness of representation-constrained AEs, we have shown an improvement to the channel charting (CC) framework in \cite{cc_paper}, where we use side information on user motion and anchor vectors to improve the positioning performance of CC. 
Numerical results for this application have shown that the use of representation constraints that are readily available in wireless positioning scenarios yield (often significant) improvements in recovered global geometry.

There are many opportunities for future work. 
While our methods enable approximate positioning, AEs with representation constraints are still not sufficient to enable GNSS-grade accuracy. Promising research directions towards this goal are the development of improved CSI features as well as the inclusion of additional geometric constraints, e.g., when acquiring CSI from multiple cell-towers or access points. 

 \balance
\bibliographystyle{IEEEtran}
\bibliography{iclr2019_conference}
\balance

\end{document}

%% file: vmr-symbols-vecbold.tex
\usepackage{amssymb}
\usepackage{amsfonts}
\usepackage{mathrsfs}
\usepackage{xspace}
\usepackage{bm}
\usepackage{upgreek}

\newcommand{\safemath}[2]{\newcommand{#1}{\ensuremath{#2}\xspace}}

\safemath{\bma}{\mathbf{a}}
\safemath{\bmb}{\mathbf{b}}
\safemath{\bmc}{\mathbf{c}}
\safemath{\bmd}{\mathbf{d}}
\safemath{\bme}{\mathbf{e}}
\safemath{\bmf}{\mathbf{f}}
\safemath{\bmg}{\mathbf{g}}
\safemath{\bmh}{\mathbf{h}}
\safemath{\bmi}{\mathbf{i}}
\safemath{\bmj}{\mathbf{j}}
\safemath{\bmk}{\mathbf{k}}
\safemath{\bml}{\mathbf{l}}
\safemath{\bmm}{\mathbf{m}}
\safemath{\bmn}{\mathbf{n}}
\safemath{\bmo}{\mathbf{o}}
\safemath{\bmp}{\mathbf{p}}
\safemath{\bmq}{\mathbf{q}}
\safemath{\bmr}{\mathbf{r}}
\safemath{\bms}{\mathbf{s}}
\safemath{\bmt}{\mathbf{t}}
\safemath{\bmu}{\mathbf{u}}
\safemath{\bmv}{\mathbf{v}}
\safemath{\bmw}{\mathbf{w}}
\safemath{\bmx}{\mathbf{x}}
\safemath{\bmy}{\mathbf{y}}
\safemath{\bmz}{\mathbf{z}}
\safemath{\bmzero}{\mathbf{0}}
\safemath{\bmone}{\mathbf{1}}

\bmdefine{\biad}{a}
\bmdefine{\bibd}{b}
\bmdefine{\bicd}{c}
\bmdefine{\bidd}{d}
\bmdefine{\bied}{e}
\bmdefine{\bifd}{f}
\bmdefine{\bigd}{g}
\bmdefine{\bihd}{h}
\bmdefine{\biid}{i}
\bmdefine{\bijd}{j}
\bmdefine{\bikd}{k}
\bmdefine{\bild}{l}
\bmdefine{\bimd}{m}
\bmdefine{\bind}{n}
\bmdefine{\biod}{o}
\bmdefine{\bipd}{p}
\bmdefine{\biqd}{q}
\bmdefine{\bird}{r}
\bmdefine{\bisd}{s}
\bmdefine{\bitd}{t}
\bmdefine{\biud}{u}
\bmdefine{\bivd}{v}
\bmdefine{\biwd}{w}
\bmdefine{\bixd}{x}
\bmdefine{\biyd}{y}
\bmdefine{\bizd}{z}

\bmdefine{\bixid}{\xi}
\bmdefine{\bilambdad}{\lambda}
\bmdefine{\bimud}{\mu}
\bmdefine{\bithetad}{\theta}
\bmdefine{\biphid}{\phi}
\bmdefine{\bideltad}{\delta}

\safemath{\bmia}{\biad}
\safemath{\bmib}{\bibd}
\safemath{\bmic}{\bicd}
\safemath{\bmid}{\bidd}
\safemath{\bmie}{\bied}
\safemath{\bmif}{\bifd}
\safemath{\bmig}{\bigd}
\safemath{\bmih}{\bihd}
\safemath{\bmii}{\biid}
\safemath{\bmij}{\bijd}
\safemath{\bmik}{\bikd}
\safemath{\bmil}{\bild}
\safemath{\bmim}{\bimd}
\safemath{\bmin}{\bind}
\safemath{\bmio}{\biod}
\safemath{\bmip}{\bipd}
\safemath{\bmiq}{\biqd}
\safemath{\bmir}{\bird}
\safemath{\bmis}{\bisd}
\safemath{\bmit}{\bitd}
\safemath{\bmiu}{\biud}
\safemath{\bmiv}{\bivd}
\safemath{\bmiw}{\biwd}
\safemath{\bmix}{\bixd}
\safemath{\bmiy}{\biyd}
\safemath{\bmiz}{\bizd}

\safemath{\bmxi}{\bixid}
\safemath{\bmlambda}{\bilambdad}
\safemath{\bmmu}{\bimud}
\safemath{\bmtheta}{\bithetad}
\safemath{\bmphi}{\biphid}
\safemath{\bmdelta}{\bideltad}

\safemath{\bA}{\mathbf{A}}
\safemath{\bB}{\mathbf{B}}
\safemath{\bC}{\mathbf{C}}
\safemath{\bD}{\mathbf{D}}
\safemath{\bE}{\mathbf{E}}
\safemath{\bF}{\mathbf{F}}
\safemath{\bG}{\mathbf{G}}
\safemath{\bH}{\mathbf{H}}
\safemath{\bI}{\mathbf{I}}
\safemath{\bJ}{\mathbf{J}}
\safemath{\bK}{\mathbf{K}}
\safemath{\bL}{\mathbf{L}}
\safemath{\bM}{\mathbf{M}}
\safemath{\bN}{\mathbf{N}}
\safemath{\bO}{\mathbf{O}}
\safemath{\bP}{\mathbf{P}}
\safemath{\bQ}{\mathbf{Q}}
\safemath{\bR}{\mathbf{R}}
\safemath{\bS}{\mathbf{S}}
\safemath{\bT}{\mathbf{T}}
\safemath{\bU}{\mathbf{U}}
\safemath{\bV}{\mathbf{V}}
\safemath{\bW}{\mathbf{W}}
\safemath{\bX}{\mathbf{X}}
\safemath{\bY}{\mathbf{Y}}
\safemath{\bZ}{\mathbf{Z}}

\safemath{\bZero}{\mathbf{0}}
\safemath{\bOne}{\mathbf{1}}
\safemath{\bDelta}{\mathbf{\Delta}}
\safemath{\bLambda}{\mathbf{\UpLambda}}
\safemath{\bPhi}{\mathbf{\Upphi}}
\safemath{\bSigma}{\mathbf{\Upsigma}}
\safemath{\bOmega}{\mathbf{\Upomega}}
\safemath{\bTheta}{\mathbf{\Uptheta}}

\bmdefine{\biAd}{A}
\bmdefine{\biBd}{B}
\bmdefine{\biCd}{C}
\bmdefine{\biDd}{D}
\bmdefine{\biEd}{E}
\bmdefine{\biFd}{F}
\bmdefine{\biGd}{G}
\bmdefine{\biHd}{H}
\bmdefine{\biId}{I}
\bmdefine{\biJd}{J}
\bmdefine{\biKd}{K}
\bmdefine{\biLd}{L}
\bmdefine{\biMd}{M}
\bmdefine{\biOd}{N}
\bmdefine{\biPd}{O}
\bmdefine{\biQd}{P}
\bmdefine{\biRd}{R}
\bmdefine{\biSd}{S}
\bmdefine{\biTd}{T}
\bmdefine{\biUd}{U}
\bmdefine{\biVd}{V}
\bmdefine{\biWd}{W}
\bmdefine{\biXd}{X}
\bmdefine{\biYd}{Y}
\bmdefine{\biZd}{Z}

\bmdefine{\biDelta}{\Delta}
\bmdefine{\biLambda}{\Lambda}
\bmdefine{\biPhi}{\Phi}
\bmdefine{\biSigma}{\Sigma}
\bmdefine{\biOmega}{\Omega}
\bmdefine{\biTheta}{\Theta}

\safemath{\bimA}{\biAd}
\safemath{\bimB}{\biBd}
\safemath{\bimC}{\biCd}
\safemath{\bimD}{\biDd}
\safemath{\bimE}{\biEd}
\safemath{\bimF}{\biFd}
\safemath{\bimG}{\biGd}
\safemath{\bimH}{\biHd}
\safemath{\bimI}{\biId}
\safemath{\bimJ}{\biJd}
\safemath{\bimK}{\biKd}
\safemath{\bimL}{\biLd}
\safemath{\bimM}{\biMd}
\safemath{\bimN}{\biNd}
\safemath{\bimO}{\biOd}
\safemath{\bimP}{\biPd}
\safemath{\bimQ}{\biQd}
\safemath{\bimR}{\biRd}
\safemath{\bimS}{\biSd}
\safemath{\bimT}{\biTd}
\safemath{\bimU}{\biUd}
\safemath{\bimV}{\biVd}
\safemath{\bimW}{\biWd}
\safemath{\bimX}{\biXd}
\safemath{\bimY}{\biYd}
\safemath{\bimZ}{\biZd}

\safemath{\bimDelta}{\biDelta}
\safemath{\bimLambda}{\biLambda}
\safemath{\bimPhi}{\biPhi}
\safemath{\bimSigma}{\biSigma}
\safemath{\bimOmega}{\biOmega}
\safemath{\bimTheta}{\biTheta}

\safemath{\setA}{\mathcal{A}}
\safemath{\setB}{\mathcal{B}}
\safemath{\setC}{\mathcal{C}}
\safemath{\setD}{\mathcal{D}}
\safemath{\setE}{\mathcal{E}}
\safemath{\setF}{\mathcal{F}}
\safemath{\setG}{\mathcal{G}}
\safemath{\setH}{\mathcal{H}}
\safemath{\setI}{\mathcal{I}}
\safemath{\setJ}{\mathcal{J}}
\safemath{\setK}{\mathcal{K}}
\safemath{\setL}{\mathcal{L}}
\safemath{\setM}{\mathcal{M}}
\safemath{\setN}{\mathcal{N}}
\safemath{\setO}{\mathcal{O}}
\safemath{\setP}{\mathcal{P}}
\safemath{\setQ}{\mathcal{Q}}
\safemath{\setR}{\mathcal{R}}
\safemath{\setS}{\mathcal{S}}
\safemath{\setT}{\mathcal{T}}
\safemath{\setU}{\mathcal{U}}
\safemath{\setV}{\mathcal{V}}
\safemath{\setW}{\mathcal{W}}
\safemath{\setX}{\mathcal{X}}
\safemath{\setY}{\mathcal{Y}}
\safemath{\setZ}{\mathcal{Z}}
\safemath{\emptySet}{\varnothing}

\safemath{\colA}{\mathscr{A}}
\safemath{\colB}{\mathscr{B}}
\safemath{\colC}{\mathscr{C}}
\safemath{\colD}{\mathscr{D}}
\safemath{\colE}{\mathscr{E}}
\safemath{\colF}{\mathscr{F}}
\safemath{\colG}{\mathscr{G}}
\safemath{\colH}{\mathscr{H}}
\safemath{\colI}{\mathscr{I}}
\safemath{\colJ}{\mathscr{J}}
\safemath{\colK}{\mathscr{K}}
\safemath{\colL}{\mathscr{L}}
\safemath{\colM}{\mathscr{M}}
\safemath{\colN}{\mathscr{N}}
\safemath{\colO}{\mathscr{O}}
\safemath{\colP}{\mathscr{P}}
\safemath{\colQ}{\mathscr{Q}}
\safemath{\colR}{\mathscr{R}}
\safemath{\colS}{\mathscr{S}}
\safemath{\colT}{\mathscr{T}}
\safemath{\colU}{\mathscr{U}}
\safemath{\colV}{\mathscr{V}}
\safemath{\colW}{\mathscr{W}}
\safemath{\colX}{\mathscr{X}}
\safemath{\colY}{\mathscr{Y}}
\safemath{\colZ}{\mathscr{Z}}

\safemath{\opA}{\mathbb{A}}
\safemath{\opB}{\mathbb{B}}
\safemath{\opC}{\mathbb{C}}
\safemath{\opD}{\mathbb{D}}
\safemath{\opE}{\mathbb{E}}
\safemath{\opF}{\mathbb{F}}
\safemath{\opG}{\mathbb{G}}
\safemath{\opH}{\mathbb{H}}
\safemath{\opI}{\mathbb{I}}
\safemath{\opJ}{\mathbb{J}}
\safemath{\opK}{\mathbb{K}}
\safemath{\opL}{\mathbb{L}}
\safemath{\opM}{\mathbb{M}}
\safemath{\opN}{\mathbb{N}}
\safemath{\opO}{\mathbb{O}}
\safemath{\opP}{\mathbb{P}}
\safemath{\opQ}{\mathbb{Q}}
\safemath{\opR}{\mathbb{R}}
\safemath{\opS}{\mathbb{S}}
\safemath{\opT}{\mathbb{T}}
\safemath{\opU}{\mathbb{U}}
\safemath{\opV}{\mathbb{V}}
\safemath{\opW}{\mathbb{W}}
\safemath{\opX}{\mathbb{X}}
\safemath{\opY}{\mathbb{Y}}
\safemath{\opZ}{\mathbb{Z}}
\safemath{\opZero}{\mathbb{O}}
\safemath{\identityop}{\opI}

\safemath{\veca}{\bma}
\safemath{\vecb}{\bmb}
\safemath{\vecc}{\bmc}
\safemath{\vecd}{\bmd}
\safemath{\vece}{\bme}
\safemath{\vecf}{\bmf}
\safemath{\vecg}{\bmg}
\safemath{\vech}{\bmh}
\safemath{\veci}{\bmi}
\safemath{\vecj}{\bmj}
\safemath{\veck}{\bmk}
\safemath{\vecl}{\bml}
\safemath{\vecm}{\bmm}
\safemath{\vecn}{\bmn}
\safemath{\veco}{\bmo}
\safemath{\vecp}{\bmp}
\safemath{\vecq}{\bmq}
\safemath{\vecr}{\bmr}
\safemath{\vecs}{\bms}
\safemath{\vect}{\bmt}
\safemath{\vecu}{\bmu}
\safemath{\vecv}{\bmv}
\safemath{\vecw}{\bmw}
\safemath{\vecx}{\bmx}
\safemath{\vecy}{\bmy}
\safemath{\vecz}{\bmz}

\safemath{\veczero}{\bmzero}
\safemath{\vecone}{\bmone}
\safemath{\vecxi}{\bmxi}
\safemath{\veclambda}{\bmlambda}
\safemath{\vecmu}{\bmmu}
\safemath{\vectheta}{\bmtheta}
\safemath{\vecphi}{\bmphi}
\safemath{\vecdelta}{\bmdelta}

\safemath{\matA}{\bA}
\safemath{\matB}{\bB}
\safemath{\matC}{\bC}
\safemath{\matD}{\bD}
\safemath{\matE}{\bE}
\safemath{\matF}{\bF}
\safemath{\matG}{\bG}
\safemath{\matH}{\bH}
\safemath{\matI}{\bI}
\safemath{\matJ}{\bJ}
\safemath{\matK}{\bK}
\safemath{\matL}{\bL}
\safemath{\matM}{\bM}
\safemath{\matN}{\bN}
\safemath{\matO}{\bO}
\safemath{\matP}{\bP}
\safemath{\matQ}{\bQ}
\safemath{\matR}{\bR}
\safemath{\matS}{\bS}
\safemath{\matT}{\bT}
\safemath{\matU}{\bU}
\safemath{\matV}{\bV}
\safemath{\matW}{\bW}
\safemath{\matX}{\bX}
\safemath{\matY}{\bY}
\safemath{\matZ}{\bZ}
\safemath{\matzero}{\bmzero}

\safemath{\matDelta}{\bDelta}
\safemath{\matLambda}{\bLambda}
\safemath{\matPhi}{\bPhi}
\safemath{\matSigma}{\bSigma}
\safemath{\matOmega}{\bOmega}
\safemath{\matTheta}{\bTheta}

\safemath{\matidentity}{\matI}
\safemath{\matone}{\matO}

\safemath{\rnda}{A}
\safemath{\rndb}{B}
\safemath{\rndc}{C}
\safemath{\rndd}{D}
\safemath{\rnde}{E}
\safemath{\rndf}{F}
\safemath{\rndg}{G}
\safemath{\rndh}{H}
\safemath{\rndi}{I}
\safemath{\rndj}{J}
\safemath{\rndk}{K}
\safemath{\rndl}{L}
\safemath{\rndm}{M}
\safemath{\rndn}{N}
\safemath{\rndo}{O}
\safemath{\rndp}{P}
\safemath{\rndq}{Q}
\safemath{\rndr}{R}
\safemath{\rnds}{S}
\safemath{\rndt}{T}
\safemath{\rndu}{U}
\safemath{\rndv}{V}
\safemath{\rndw}{W}
\safemath{\rndx}{X}
\safemath{\rndy}{Y}
\safemath{\rndz}{Z}

\safemath{\rveca}{\bimA}
\safemath{\rvecb}{\bimB}
\safemath{\rvecc}{\bimC}
\safemath{\rvecd}{\bimD}
\safemath{\rvece}{\bimE}
\safemath{\rvecf}{\bimF}
\safemath{\rvecg}{\bimG}
\safemath{\rvech}{\bimH}
\safemath{\rveci}{\bimI}
\safemath{\rvecj}{\bimJ}
\safemath{\rveck}{\bimK}
\safemath{\rvecl}{\bimL}
\safemath{\rvecm}{\bimM}
\safemath{\rvecn}{\bimN}
\safemath{\rveco}{\bomO}
\safemath{\rvecp}{\bimP}
\safemath{\rvecq}{\bimQ}
\safemath{\rvecr}{\bimR}
\safemath{\rvecs}{\bimS}
\safemath{\rvect}{\bimT}
\safemath{\rvecu}{\bimU}
\safemath{\rvecv}{\bimV}
\safemath{\rvecw}{\bimW}
\safemath{\rvecx}{\bimX}
\safemath{\rvecy}{\bimY}
\safemath{\rvecz}{\bimZ}

\safemath{\rvecxi}{\bmxi}
\safemath{\rveclambda}{\bmlambda}
\safemath{\rvecmu}{\bmmu}
\safemath{\rvectheta}{\bmtheta}
\safemath{\rvecphi}{\bmphi}

\safemath{\rmatA}{\bimA}
\safemath{\rmatB}{\bimB}
\safemath{\rmatC}{\bimC}
\safemath{\rmatD}{\bimD}
\safemath{\rmatE}{\bimE}
\safemath{\rmatF}{\bimF}
\safemath{\rmatG}{\bimG}
\safemath{\rmatH}{\bimH}
\safemath{\rmatI}{\bimI}
\safemath{\rmatJ}{\bimJ}
\safemath{\rmatK}{\bimK}
\safemath{\rmatL}{\bimL}
\safemath{\rmatM}{\bimM}
\safemath{\rmatN}{\bimN}
\safemath{\rmatO}{\bimO}
\safemath{\rmatP}{\bimP}
\safemath{\rmatQ}{\bimQ}
\safemath{\rmatR}{\bimR}
\safemath{\rmatS}{\bimS}
\safemath{\rmatT}{\bimT}
\safemath{\rmatU}{\bimU}
\safemath{\rmatV}{\bimV}
\safemath{\rmatW}{\bimW}
\safemath{\rmatX}{\bimX}
\safemath{\rmatY}{\bimY}
\safemath{\rmatZ}{\bimZ}

\safemath{\rmatDelta}{\bimDelta}
\safemath{\rmatLambda}{\bimLambda}
\safemath{\rmatPhi}{\bimPhi}
\safemath{\rmatSigma}{\bimSigma}
\safemath{\rmatOmega}{\bimOmega}
\safemath{\rmatTheta}{\bimTheta}

%% file: standard-macros.tex
\usepackage{amssymb}
\usepackage{amsfonts}
\usepackage{mathrsfs}
\usepackage{xspace}
\usepackage{bm}
\usepackage{fancyref}
\usepackage{textcomp}

\usepackage{multirow}
\usepackage{stmaryrd}

\newenvironment{textbmatrix}{	\setlength{\arraycolsep}{2.5pt}%
								\big[\begin{matrix}}{\end{matrix}\big]%
								\raisebox{0.08ex}{\vphantom{M}}}

\def\be{\begin{equation}}
\def\ee{\end{equation}}
\def\een{\nonumber \end{equation}}
\def\mat{\begin{bmatrix}}
\def\emat{\end{bmatrix}}
\def\btm{\begin{textbmatrix}}
\def\etm{\end{textbmatrix}}

\def\ba#1\ea{\begin{align}#1\end{align}}
\def\bas#1\eas{\begin{align*}#1\end{align*}}
\def\bs#1\es{\begin{split}#1\end{split}} 
\def\bg#1\eg{\begin{gather}#1\end{gather}}
\def\bml#1\eml{\begin{multline}#1\end{multline}}
\def\bi#1\ei{\begin{itemize}#1\end{itemize}}

\safemath{\dirac}{\delta}					% Dirac delta
\safemath{\krond}{\dirac}					% Kronecker delta
\safemath{\upto}{\uparrow}
\safemath{\downto}{\downarrow}
\safemath{\iu}{j}							% imaginary unit
\safemath{\ev}{\lambda}						% eigenvalue
\safemath{\hilseqspace}{l^{2}}				% Hilbert sequence space
\newcommand{\banachfunspace}[1]{\setL^{#1}}	% Banach function space
\safemath{\hilfunspace}{\banachfunspace{2}}	% Hilbert function space
\safemath{\SNR}{\textsf{SNR}} 				% signal to noise ratio
\safemath{\PAR}{\textsf{PAR}} 				% signal to noise ratio
\safemath{\No}{N_0}							% noise spectral density
\safemath{\Es}{E_s}							% energy per symbol
\safemath{\Eb}{E_b}							% energy per bit
\safemath{\EbNo}{\frac{\Eb}{\No}}
\safemath{\EsNo}{\frac{\Es}{\No}}

\DeclareMathOperator{\CHop}{\ensuremath{\opH}} % channel operator
\safemath{\tvir}{\rndh_{\CHop}}				% time-varying impulse response
\safemath{\tvtf}{\rndl_{\CHop}}				% 	-''- transfer function
\safemath{\spf}{\rnds_{\CHop}}				% spreading function
\safemath{\bff}{H_{\CHop}}					% bi-freuqency function

\safemath{\ircf}{r_{h}}						% impulse response correlation fn.
\safemath{\tftvcf}{r_{s}}					% scattering function
\safemath{\tfcf}{r_{l}}						% time-frequency correlation fn.
\safemath{\bfcf}{r_{H}}						% bi-frequency correlation fn.

\safemath{\tcorr}{c_h}						% time-correlation function
\safemath{\scf}{c_{s}}						% spreading function
\safemath{\tfcorr}{c_{l}}					% transfer-function correlation
\safemath{\fcorr}{c_{H}}						% frequency-correlation function

\safemath{\mi}{I}							% mutual information
\safemath{\capacity}{C}						% capacity

\safemath{\normal}{\mathcal{N}}			% normal distribution
\safemath{\jpg}{\mathcal{CN}}			% jointly proper Gaussian
\safemath{\mchain}{\leftrightarrow}		% Markov chain
\safemath{\dB}{\,\mathrm{dB}}
\safemath{\dBm}{\,\mathrm{dBm}}
\safemath{\Hz}{\,\mathrm{Hz}}
\safemath{\kHz}{\,\mathrm{kHz}}
\safemath{\MHz}{\,\mathrm{MHz}}
\safemath{\GHz}{\,\mathrm{GHz}}
\safemath{\s}{\,\mathrm{s}}
\safemath{\ms}{\,\mathrm{ms}}
\safemath{\mus}{\,\mathrm{\text{\textmu}s}}
\safemath{\ns}{\,\mathrm{ns}}
\safemath{\ps}{\,\mathrm{ps}}
\safemath{\meter}{\,\mathrm{m}}
\safemath{\mm}{\,\mathrm{mm}}
\safemath{\cm}{\,\mathrm{cm}}
\safemath{\m}{\,\mathrm{m}}
\safemath{\W}{\,\mathrm{W}}
\safemath{\mW}{\, \mathrm{mW}}
\safemath{\J}{\,\mathrm{J}}
\safemath{\K}{\,\mathrm{K}}
\safemath{\bit}{\,\mathrm{bit}}
\safemath{\nat}{\,\mathrm{nat}}

\safemath{\define}{\triangleq}			% definition

\safemath{\equivalent}{\sim}
\safemath{\distas}{\sim}					% distributed according to
\safemath{\sdiff}{\Delta}				% symmetric set difference

\safemath{\reals}{\mathbb{R}}
\safemath{\positivereals}{\reals_{+}}
\safemath{\integers}{\mathbb{Z}}
\safemath{\posint}{\integers_{+}}
\safemath{\naturals}{\mathbb{N}}
\safemath{\posnaturals}{\naturals_{+}}
\safemath{\complexset}{\mathbb{C}}
\safemath{\rationals}{\mathbb{Q}}

\newcommand*{\fancyrefapplabelprefix}{app}		% Appendix
\newcommand*{\fancyrefthmlabelprefix}{thm}		% Theorem
\newcommand*{\fancyreflemlabelprefix}{lem}		% Lemma
\newcommand*{\fancyrefcorlabelprefix}{cor}		% Corollary
\newcommand*{\fancyrefdeflabelprefix}{def}		% Definition
\newcommand*{\fancyrefproplabelprefix}{prop}	% Proposition
\newcommand*{\fancyrefobslabelprefix}{obs}		% Observation 
\newcommand*{\fancyrefalglabelprefix}{alg}		% Algorithm
\newcommand*{\fancyrefasmlabelprefix}{asm}	    % Assumption
\newcommand*{\fancyrefasmslabelprefix}{asms}	    % Assumptions
\newcommand*{\fancyreftbllabelprefix}{tbl}	    % Assumption
\newcommand*{\fancyreftremabelprefix}{rem}	    % Remark

\frefformat{vario}{\fancyrefseclabelprefix}{Section~#1}
\frefformat{vario}{\fancyrefthmlabelprefix}{Theorem~#1}
\frefformat{vario}{\fancyreflemlabelprefix}{Lemma~#1}
\frefformat{vario}{\fancyrefcorlabelprefix}{Corollary~#1}
\frefformat{vario}{\fancyrefdeflabelprefix}{Definition~#1}
\frefformat{vario}{\fancyrefobslabelprefix}{Observation~#1}
\frefformat{vario}{\fancyrefasmlabelprefix}{Assumption~#1}
\frefformat{vario}{\fancyrefasmslabelprefix}{Assumptions~#1}
\frefformat{vario}{\fancyreffiglabelprefix}{Figure~#1}
\frefformat{vario}{\fancyrefapplabelprefix}{Appendix~#1} 
\frefformat{vario}{\fancyrefproplabelprefix}{Proposition~#1}
\frefformat{vario}{\fancyrefalglabelprefix}{Algorithm~#1}
\frefformat{vario}{\fancyrefeqlabelprefix}{(#1)}
\frefformat{vario}{\fancyreftbllabelprefix}{Table~#1}
\frefformat{vario}{\fancyreftremabelprefix}{Remark~#1}

%% file: defs.tex
\safemath{\dictab}{[\,\dicta\,\,\dictb\,]}

\safemath{\ysig}{\bmy}
\safemath{\ysighat}{\hat{\ysig}}
\safemath{\ysigdim}{M}
\safemath{\xsig}{\bmx}
\safemath{\xsigdim}{N}
\safemath{\nx}{n_x}
\safemath{\zsig}{\bmz}
\safemath{\zsigdim}{\ysigdim}
\safemath{\rsig}{\bmr}
\safemath{\Adict}{\bA}
\safemath{\Adicttilde}{\widetilde{\Adict}}
\safemath{\Adictdim}{\outputdim\times\xsigdim}
\safemath{\avec}{\bma}
\safemath{\avectilde}{\tilde{\avec}}
\safemath{\Bdict}{\bB}
\safemath{\Bdicttilde}{\widetilde{\Bdict}}
\safemath{\Cdict}{\bC}
\safemath{\cvec}{\bmc}
\safemath{\Ddict}{\bD}
\safemath{\Ddictdim}{\ysigdim\times\xsigdim}
\safemath{\dvec}{\bmd}
\safemath{\Ddicttilde}{\widetilde{\bD}}
\safemath{\Bonb}{\bB}
\safemath{\bvec}{\bmb}
\safemath{\Bonbdim}{\ysigdim\times\ysigdim}
\safemath{\noise}{\bmn}
\safemath{\noisedim}{\ysigim}
\safemath{\err}{\bme}
\safemath{\errdim}{\ysigdim}
\safemath{\errset}{\setE}
\safemath{\nerr}{n_e}
\safemath{\delop}{\bP_\errset}
\safemath{\delopc}{\bP_{{\errset}^c}}

\safemath{\cplxi}{\imath}
\safemath{\cplxj}{\jmath}
\safemath{\dict}{\matD}
\safemath{\inputdim}{N}		% number of columns of dictionary D
\safemath{\outputdim}{M}		%number of rows of dictionary D
\safemath{\sparsity}{S}	%sparsity
\safemath{\inputdimA}{{N_a}}	%total number of elements in dictionary A
\safemath{\inputdimB}{{N_b}}	%total number of elements in dictionary B
\safemath{\elemA}{{n_a}}	%number of elements chosen from dictionary A
\safemath{\elemB}{{n_b}}	%number of elements chosen from dictionary B
\safemath{\resA}{\matR_a}	%restriction map to elements of dictionary A
\safemath{\resB}{\matR_b}	%restriction map to elements of dictionary B
\safemath{\subD}{\matS} %subdictionary
\safemath{\subA}{\matS_a} %subdictionary part of A
\safemath{\subB}{\matS_b} %subdictionary part of B
\safemath{\dicta}{\matA} 	% first subdictionary
\safemath{\dictb}{\matB} 	% second subdictionary
\safemath{\hollowS}{H}
\safemath{\hollowA}{H_a}
\safemath{\hollowB}{H_b}
\safemath{\cross}{Z}
\safemath{\coh}{\mu_d}			% coherence dictionary
\safemath{\coha}{\mu_a}			% coherence first subdictionary
\safemath{\cohb}{\mu_b}			% coherence second subdictionary
\safemath{\mubs}{\nu}	%block sub-coherence
\safemath{\cohm}{\mu_m} %mutual coherence
\safemath{\dictset}{\setD}	% set of dictionaries
\safemath{\dictsetp}{\dictset(\coh,\coha,\cohb)}	% set of dictionaries parametrized
\safemath{\dictsetgen}{\dictset_\text{gen}}
\safemath{\dictsetgenp}{\dictsetgen(\coh)}
\safemath{\dictsetonb}{\dictset_\text{onb}}
\safemath{\dictsetonbp}{\dictsetonb(\coh)}

\safemath{\leftside}{U}
\safemath{\rightsideA}{R_a}
\safemath{\rightsideB}{R_b}

\safemath{\indexS}{\setI_S} %set of indices participating in sub-dictionary S

\safemath{\na}{n_a}			% cardinality of set of linearly independent columns of first dictionary
\safemath{\nb}{n_b}			% cardinality of set of linearly independent columns of second dictionary
\safemath{\coeffa}{p_i}	%coefficients for columns of A
\safemath{\coeffb}{q_j}	%coefficients for columns of B
\safemath{\seta}{\setP}		% set of linearly independent columns of A
\safemath{\setb}{\setQ}     % set of linearly independent columns of B
\safemath{\setw}{\setW}	%set of n largest elements of w
\safemath{\setz}{\setZ}	%set of L-n largest elements of z
\safemath{\cola}{\veca}		% generic element of the dictionary A
\safemath{\colb}{\vecb}		% generic element of the dictionary B
\safemath{\cold}{\vecd}		% generic element of the dictionary D
\safemath{\inputvec}{\vecx} 	%coefficient vector (input)
\safemath{\error}{\vece}	%error vector
\safemath{\noiseout}{\vecz} 	%noisy output vector
\safemath{\inputvecel}{x}
\safemath{\inputveca}{\vecx_a}
\safemath{\inputvecb}{\vecx_b}
\safemath{\outputvec}{\vecy}	%output of Dictionary
\safemath{\lambdamin}{\lambda_{\mathrm{min}}}
\safemath{\elltwo}{\ell_2}
\safemath{\ellone}{\ell_1}
\safemath{\ellzero}{\ell_0}
\safemath{\ellinf}{\ell_\infty}
\safemath{\ellinftilde}{\ell_{\widetilde\infty}}
\safemath{\licard}{Z(\coh,\coha,\cohb)}
\safemath{\xsol}{\hat{x}}
\safemath{\xbord}{x_b}		%Solution at the border
\safemath{\xstat}{x_s}		%Solution stationary in l0 prob
\safemath{\xstatLone}{\tilde{x}_s}
\safemath{\order}{\mathcal{O}} %order notation (big O)
\safemath{\scales}{\Theta} %scales as
\safemath{\ones}{\mathbf{1}} %all ones matrix
\safemath{\zeroes}{\mathbf{0}} %all zeroes matrix
\safemath{\thlone}{\kappa(\coh,\cohb)} %treshold l1 problem
\safemath{\constoneA}{\delta} %constant in l1 theorem to save space
\safemath{\constoneB}{\epsilon} %constant in l1 theorem to save space
\safemath{\nlarge}{L}				   %num large elements
\safemath{\sumlarge}{S_\nlarge}
\safemath{\maxlarger}{P_\nlarge}	   % maximum in Gribonval and Nielsen
\safemath{\Pzero}{\textrm{P0}}	
\safemath{\Pone}{\textrm{P1}}
\safemath{\vecfir}{\vecw}			 % \vecv element of the kernel of the dictionary, \vecv=[\vecfir \vecsec]
\safemath{\vecsec}{\vecz}
\safemath{\elvecfir}{w}              % element of vecfir
\safemath{\elvecsec}{z}				 % element of vecsec
\safemath{\nlargefir}{n}
\safemath{\normout}{\gamma}
\safemath{\auxfun}{h}
\safemath{\supp}{\textrm{supp}}%support

\safemath{\indexa}{\ell}
\safemath{\indexb}{r}
\safemath{\indexc}{i}
\safemath{\indexd}{j}

\safemath{\project}{P}%projector